\documentclass[twocolumn,aps,prl,showpacs,floatfix]{revtex4}
\usepackage{amsmath,bm}
\usepackage{graphicx}
\usepackage{epsfig}
\begin{document}

\title{Shear viscosity to entropy density ratio in nuclear multifragmentation}
\author{Subrata Pal}
\affiliation{Department of Nuclear and Atomic Physics, Tata Institute of
Fundamental Research, Homi Bhabha Road, Mumbai 400005, India} 

\begin{abstract} 
Nuclear multifragmentation in intermediate energy heavy ion collisions
has long been associated with liquid-gas phase transition. We calculate
the shear viscosity to entropy density ratio $\eta/s$ for an equilibrated 
system of nucleons and fragments produced in multifragmentation within 
an extended statistical multifragmentation model. The temperature dependence of 
$\eta/s$ exhibits surprisingly similar behavior as that for H$_2$O. In the 
coexistence phase of fragments and light particles, the ratio $\eta/s$ reaches a 
minimum of comparable depth as that for water in the vicinity of the critical 
temperature for liquid-gas phase transition. The effects of freeze-out volume 
and surface symmetry energy on $\eta/s$ in multifragmentation are studied.
\end{abstract}

\maketitle

Understanding the behavior of nuclear matter under extreme conditions has been one 
of the most important present challenges in heavy ion physics. Multifragmentation 
in intermediate energy heavy ion collisions provide a key mechanism to 
address this issue where an excited nucleus formed expands and breaks up into 
various fragments and light particles \cite{Gross,Bondorf,Das}. 
The final yield distribution is quite sensitive to the internal excitation, breakup 
density of the nucleus and the symmetry energy part of the binding energy of the 
fragments \cite{Botvina,Fevre,Shetty76,Tsang92,Souza78}. 

Due to van der Waals nature
of the nucleon-nucleon interaction, it is expected that multifragmentation may exhibit
features of liquid-gas phase transition \cite{Gross,Bondorf}. Evidence of this provided from 
the observation of the nuclear caloric curve relating excitation energy to the temperature 
of the breakup source \cite{Poch,Nato,Agos}. Extensive studies have been carried out 
to understand the dependence of the caloric curve on the system size and particularly on the 
density dependence of nuclear symmetry energy which is poorly constrained \cite{DLL}.

Multifragmentation studies are however mostly confined to the effects of the state 
variables on the thermodynamic properties of the system. Whereas transport properties of the 
dynamically evolving system of nucleons and fragments formed in fragmentation have received 
little attention. Since the transport coefficients characterize the dynamics of fluctuation of the
dissipative fluxes in a medium \cite{Landau}, their knowledge is essential for a better 
understanding of the fragmentation observables. Recently, considerable attention has been 
focussed primarily on the shear viscosity coefficient that involves the transport of momentum 
due to velocity gradient in an anisotropic medium. Empirical observation \cite{Csernai} 
of the temperature dependence of the shear viscosity to entropy density ratio $\eta/s$ 
for water (as well as He, Ne$_2$) exhibits a minimum in the vicinity 
of the critical temperature $T_c$ for liquid-gas phase transition. Furthermore, a lower 
bound of $\eta/s \geq 1/4\pi$ obtained by Kovtun-Son-Starinets (KSS) \cite{KSS} in certain 
gauge theories is speculated \cite{KSS} to be valid for several substances in nature.

In heavy ion collisions at energy $E_{\rm lab} \lesssim 1000$ MeV/nucleon, the shear 
viscosity coefficient has been estimated for nucleon transport in the Uhlenbeck-Uehling 
equations \cite{Danielewicz}. Analysis of the 
observed transverse flow of nucleons in the microscopic QMD model in Au+Au, Nb+Nb and
Ca+Ca central collisions at energy $E_{\rm lab} =400-1200$ MeV/nucleons requires shear 
viscosity of $30 <\eta< 60$ MeV/(fm$^2$c) \cite{Schur}. Since the degrees of freedom in 
these studies involve only nucleons hence the $\eta$ estimate is relevant only in the early 
stages of the reaction. Whereas fragments formed due to correlations and fluctuations of the 
system that goes beyond the mean field dynamics of nucleon transport should result in a different
shear viscosity. At freeze-out the momentum transport and thereby the shear viscosity is 
mostly effected by long range Coulomb interaction between the charged particles. 

In this paper we estimate the shear viscosity to entropy density ratio $\eta/s$ of
an equilibrated system of fragments and/or light particles produced from multifragmentation
within a modified microcanonical statistical multifragmentation model (SMM) \cite{Bondorf} 
that has successfully reproduced several observables \cite{Bondorf,Das,Botvina,Souza76}.
We shall show that $\eta/s$ decreases as a function of temperature of 
the system, exhibits a minimum at the coexisting liquid-gas phase comprising of 
intermediate mass fragments and light particles, and increases again at high temperature
in a system of light particles. The estimated minimum of $\eta/s$ is found 
comparable to that for water in the vicinity of critical point for liquid-gas phase transition.

In the SMM a hot source with mass and charge ($A_0,Z_0$) at temperature $T$ expands to a
freeze-out volume $V=(1+\chi)V_0$ and undergoes prompt statistical breakup; here $V_0$ is
the normal nuclear volume and $\chi \geq 0$ is input parameter. We start with the
grand-canonical version of SMM which consists of minimizing the free energy $F$ of the
system:
\begin{equation} \label{free}
F(T) = \sum_{A,Z} N_{A,Z} \left[ f^{\rm tr}_{A,Z} + f^*_{A,Z} \right] + F_c ~,
\end{equation}
under conservation of mass $A_0=\sum_{A,Z}N_{A,Z}A$ and charge $Z_0=\sum_{A,Z}N_{A,Z}Z$.
The translational energy of a nuclear species with mass and charge ($A,Z$) is
\begin{equation} \label{ftrn}
f^{\rm tr}_{A,Z} = -T \left[ \log\left(\frac{g_{A,Z} V_f A^{3/2}}{\lambda_T^3} \right)
- \frac{\log(N_{A,Z}!)}{N_{A,Z}} \right] ~,
\end{equation}
where the thermal wavelength is $\lambda_T = \sqrt{2\pi\hbar^2/m T}$, $m$ is the
nucleon mass. The spin degeneracy is denoted by $g_{A,Z}$ and the free volume $V_f = V-V_0$. 
The multiplicity $N_{A,Z}$ of the fragment ($A,Z$) is then given by 
\cite{Bondorf,Souza78}
\begin{eqnarray} \label{GCM}
N_{A,Z} &=& \frac{g_{A,Z} V_f A^{3/2}}{\lambda_T^3} \exp \Big\{ -\Big[ 
f^*_{A,Z} - \mu_BA -\mu_QZ \nonumber\\
&& + \frac{2 C_c Z Z_0}{(1+\chi)^{1/3} A_0^{1/3}}
- \frac{C_c A Z_0^2}{3(1+\chi)^{1/3} A_0^{4/3}}  \Big] \Big/ T \Big\} ~.
\end{eqnarray}
The baryon and charge chemical potentials $\mu_B$ and $\mu_Q$ are obtained from mass 
and charge conservation. 
The internal free energy of the species ($A,Z$) is
\begin{eqnarray} \label{fint}
f^*_{A,Z} &=& - B_{A,Z} - \frac{T^2}{\epsilon_0}A 
+ \beta_0 A^{2/3}\left[ \left(\frac{T_c^2-T^2}{T_c^2+T^2}\right)^{5/4} -1 \right] \nonumber\\
&& - \frac{C_c Z^2}{(1+\chi)^{1/3} A^{1/3}} ~,
\end{eqnarray}
where the parameters are $\epsilon_0=16$ MeV, $\beta_0=18$ MeV and $T_c=18$ MeV.
The fragments here are assumed to be at normal density.
The Coulomb repulsion between the fragment $F_c$ in Eq. (\ref{free}) is evaluated
in the Wigner-Seitz approximation and corresponds to the last term in Eq. (\ref{fint}).
The Coulomb self energy of a fragment is included in its binding energy $B_{A,Z}$.
The last two terms in Eq. (\ref{GCM}) stem from homogeneous 
term of the Wigner-Seitz approximation \cite{Bondorf,Souza78}. Light nuclei with $A<5$
are considered point particles with no internal degrees of freedom so that the bulk 
and surface contributions to internal free energy $f^*_{A,Z}$ [second and third term 
of Eq. (\ref{fint}), respectively] are neglected, except for alpha particle, 
where the bulk contribution is retained as usual \cite{Bondorf}. For these
light nuclei, experimental values for binding energy are used. For heavier
nuclei $A \geq 5$, the spin degeneracy $g_{A,Z}=1$, and for binding energy $B_{A,Z}$, 
the computed Liquid Drop Mass (LDM) formula of Ref. \cite{Souza78} is used:
\begin{equation} \label{LDM}
B_{A,Z} =  C_vA - C_sA^{2/3} - C_c\frac{Z^2}{A^{1/3}} + C_d\frac{Z^2}{A} ~,
\end{equation}
where $C_i = a_i[1-k_i(A-2Z)^2/A^2]$ and $i=v,s$ corresponds to the volume and surface
terms, respectively. The volume and surface contributions to symmetry energy give
$E_{\rm sym} = C_{\rm sym}(A-2Z)^2/A^2$ where $C_{\rm sym} = a_vk_v - a_sk_s/A^{1/3}$.
Two versions of LDM formula are used here, dubbed as LDM1 and LDM2 \cite{Souza78},
to study surface symmetry energy effects on the $\eta/s$ ratio. The simpler LDM1 
has $k_s=0$ (also $C_d=0$) and thus neglects the surface corrections to the 
symmetry energy. The complete LDM2 formula preserves all the terms.

The energy conservation of the fragmenting source leads to
\begin{eqnarray} \label{energy}
&&E_{\rm sour}^{\rm gs} + E^* = E^{\rm tr}(T) + \sum_{A,Z} N_{A,Z} \left[ -B_{A,Z} 
+ \varepsilon^*_{A,Z} \right] \nonumber\\
&&+ \frac{C_c}{(1+\chi)^{1/3}} \frac{Z_0^2}{A_0^{1/3}}
- \frac{C_c}{(1+\chi)^{1/3}} \sum_{A,Z} N_{A,Z} \frac{Z^2}{A^{1/3}} ~,
\end{eqnarray}
where $E_{\rm sour}^{\rm gs}$ is the ground state energy of the source and
$\varepsilon^*_{A,Z}(T)$ denotes the excitation energy of the fragment ($A,Z$)
at temperature $T$ with a translational energy $\varepsilon^{\rm tr}_{A,Z} =3T/2$.
Equation (\ref{energy}) allows to extract the excitation energy $E^*$ of the
source at a given $T$ and thereby construct the nuclear caloric curve.

In the grand-canonical SMM the probability distribution of the fragment yield, 
$P_{A,Z} = N_{A,Z}/ \sum_{A,Z}N_{A,Z}$, allows one to generate \cite{PalMC} 
a Monte Carlo microcanonical ensemble of fragments with exact conservation of 
mass $A_0$, charge $Z_0$ and total energy of the source. The fragments in a microcanonical
ensemble are placed in a non-overlapping fashion within a spherical freeze-out
volume $V$. These particle are then allowed to evolve in time under Coulomb repulsion 
within the freeze-out volume with periodic boundary conditions in the configuration 
space. Finally, collision between the fragments enforce kinetic equilibration when the
system is found to approach momentum isotropization \cite{Pal}. The shear viscosity
of this dynamically evolving equilibrated system is then estimated.

\begin{figure}[ht]
\centerline{\epsfig{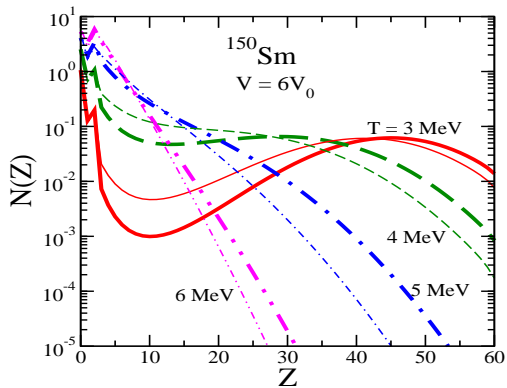}}
\vspace{-0.2cm}

\caption{(Color online) Charge distribution in the breakup of $^{150}$Sm nucleus at 
various temperatures $T$ at a freeze-out volume of $V=6V_0$ in the LDM1 (thin lines) 
and LDM2 (thick lines) mass formulas.}

\label{charge}
\end{figure}

In Fig. \ref{charge} we show the charge distribution for the breakup of $^{150}$Sm at
$T=3,4,6$ MeV in the LDM1 and LDM2 mass formulas. At low temperature $T=3$ MeV
the system is characterized by few light particles and a massive nucleus (liquid phase) 
close to the source charge. The minimum of the free energy $F=E-TS$ is 
essentially controlled by the surface term in Eq. (\ref{fint}) that favors one massive 
nucleus instead of small fragments with a larger total surface. 
With increasing temperature more intermediate mass fragments, IMFs ($3 \leq Z \leq 15$),
are produced. Here the $-TS$ term in free energy, dominated by binding energy 
and Coulomb repulsion, favors the breakup into small sized fragments \cite{Das}.
Eventually at large $T$ only light particles are produced (gas phase).
Compared to the LDM1 set, the inclusion of surface symmetry energy in LDM2
that are important for light nuclei of mass $A<25$ \cite{Souza78}, leads to the suppression
of isospin symmetric IMFs and an enhancement of neutron rich heavier nuclei.

\begin{figure}[ht]
\centerline{\epsfig{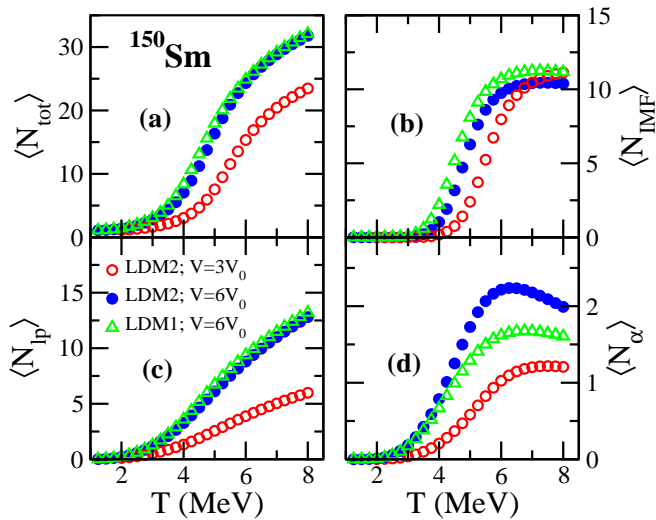}}
\vspace{-0.2cm}

\caption{(Color online) The temperature dependence of multiplicity of (a) all particles, 
$N_{\rm tot}$, (b) IMFs, $N_{\rm IMF}$, (c) light particles with $A<5$ except alpha,
$N_{\rm lp}$, and (d) alpha, $N_\alpha$, in the breakup of $^{150}$Sm nucleus in 
LDM1 (triangles) and LDM2 (solid circles) mass formula 
at a freeze-out volume of $V=6V_0$ and in LDM2 (open circles) at $V=3V_0$.}

\label{mulp}
\end{figure}

In order to get insight into the quantitative effects from the two mass formulas,
we show in Fig. \ref{mulp}, the multiplicity of different species as a function
of temperature. At low $T < 3$ MeV where the yield is dominated by a massive fragment, 
the symmetry energy effects are imperceptible in the LDM1 (triangles) and LDM2 (solid circles)
sets at the same freeze-out volume $V=6V_0$. At moderate and high temperatures the 
total multiplicity $N_{\rm tot}$ in LDM2 set is reduced as surface symmetry energy 
favors heavier nuclei with large neutron-proton asymmetry. This
leads to a reduction in the yields of more isospin symmetric IMFs, $N_{\rm IMF}$, and 
light nuclei, $N_{\rm lp}$, in comparison to the LDM1 set. Also shown in Fig. \ref{mulp}
the particle abundances in the LDM2 set at a smaller freeze-out volume $V=3V_0$,
where as expected, the yields are reduced.

\begin{figure}[ht]
\centerline{\epsfig{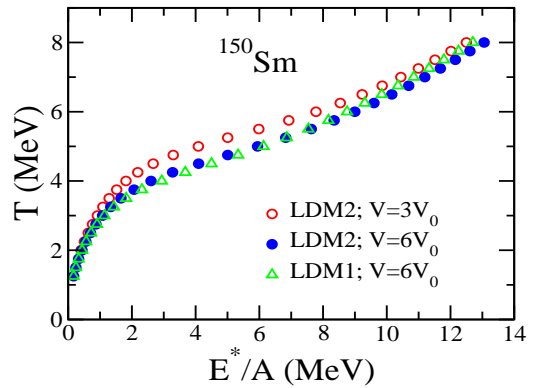}}
\vspace{-0.2cm}

\caption{(Color online) Caloric curve expressing excitation energy per nucleon
$E^*/A$  versus temperature $T$ for the breakup of $^{150}$Sm nucleus in 
LDM1 (triangles) and LDM2 (solid circles) mass formula at a freeze-out volume of 
$V=6V_0$ and in LDM2 (open circles) at $V=3V_0$.}

\label{caloric}
\end{figure}

The nuclear caloric curve \cite{Bondorf,Poch,Nato,Agos} associated with the 
fragmentation of $^{150}$Sm nucleus is 
shown in Fig. \ref{caloric}. At $3 \lesssim E^*/A \lesssim 8$ MeV, the caloric curve
exhibits a slow but monotonous increase of temperature with excitation energy
\cite{Nato,Aguiar,Botvina08} which may be a signature of liquid-gas phase transition. 
This behavior stems from energy conservation constraint of Eq. (\ref{energy}) when appreciable 
amount of energy is used to produce abundant IMFs and light particles. While the experimentally
observed plateau \cite{Nato,Agos} in the caloric curve may result if the breakup occurs
at a fixed pressure (i.e. multifragmentation is an isobaric process) \cite{Aguiar,Elliot,Chomaz} 
in contrast to the fixed freeze-out volume employed here, or if the fragments are expanded 
\cite{Souza09} unlike the fragments assumed here to be at normal nuclear density. Note in this 
excitation range, the massive fragments with large binding and internal excitation in the 
LDM2 set (solid circles) lead to slightly higher temperatures (or conversely smaller $E^*/A$) 
compared to LDM1 set (triangles). While at high $E^*/A > 8$ MeV, the smaller 
multiplicity of light particles, which have no internal degrees of freedom, 
in the LDM2 set results in lower temperatures relative to LDM1 set. 
At freeze-out volume $V=3V_0$ in LDM2, the breakup temperature is consistently higher 
as more massive and fewer fragments are produced.

\begin{figure}[ht]
\centerline{\epsfig{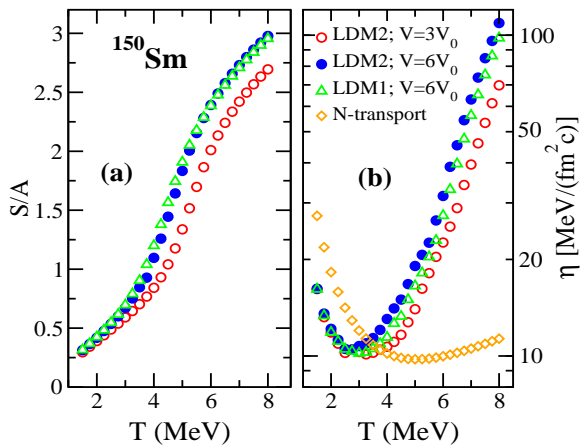}}
\vspace{-0.2cm}

\caption{(Color online) (a) Entropy per nucleon $S/A$, and (b) shear viscosity 
coefficient $\eta$ as a function of temperature for the breakup of $^{150}$Sm nucleus
in LDM1 (triangles) and LDM2 (solid circles) mass formula at a freeze-out volume of 
$V=6V_0$ and in LDM2 (open circles) at $V=3V_0$. Shear viscosity in nucleon transport
calculation \cite{Danielewicz} is shown in diamond.}

\label{visco}
\end{figure}

The entropy $S=-dF/dT$ in the microcanonical ensemble of fragments is 
determined using the conventional thermodynamic relation \cite{Bondorf,Das}
\begin{eqnarray} \label{entr}
S &=& \log\prod_{A,Z}(2g_{A,Z}+1) 
+ \log\prod_{A,Z} A^{3/2} - \ln A_0^{3/2} \nonumber\\
&& - \log\Big(\prod_{A,Z}(N_{A,Z}!\Big) + (M-1)\log(V_f/\lambda_T^3) \nonumber\\ 
&& + 1.5(M-1) -\sum_{A,Z}\partial f^*_{A,Z}/\partial T ~,
\end{eqnarray}
where the total multiplicity in an event is $M=\sum_{A,Z}N_{A,Z}$ and the last term
corresponds to entropy contribution from the bulk and surface terms in the internal
free energy of Eq. (\ref{fint}).

In Fig. \ref{visco}(a) the entropy per nucleon $S/A$ is shown as a
function of temperature for fragmentation of $^{150}$Sm. In the temperature
range $4 \leq T \leq 6$ MeV associated with liquid-gas mixed phase in the
caloric curve, the entropy shows a rapid increase with temperature. In this region
abundant IMFs are produced in the LDM1 set (triangles) with smaller internal degrees of 
freedom that leads to somewhat higher entropy than in LDM2 set (solid circles)
with more massive fragments. At $V=3V_0$ in LDM2, suppression of particle yield 
leads to a reduction in entropy.

For time evolving system of fragments in equilibrium, the shear 
viscosity due to momentum transport (via Coulomb scattering between the fragments) 
can be computed from Kubo relation \cite{Kubo,Pal} or in the  classical
kinetic theory \cite{Reif}. The Kubo formula employs the linear response 
theory to relate the transport coefficients as correlations of dissipative
fluxes. However it provides the total viscosity of the system and not
from individual species. On the other hand, in the kinetic theory, the
total shear viscosity of a multicomponent system can be expressed as the
sum from individual contribution as 
$\eta \simeq (1/3) \sum_i n_i \langle p_i\rangle \lambda_i$, where $n_i$ is the
number density, $\langle p_i\rangle$ is the average momentum and $\lambda_i$ is 
the mean free path of the $i$th species. Moreover, 
$\lambda_i = 1/\sum_j n_j\sigma_{ij}$ where $\sigma_{ij}$ is the collisional
cross section which is taken as that for the usual Coulomb scattering. 
The average thermal momentum 
of the particle in the nonrelativistic limit is 
$\langle p_i\rangle = m_i\langle v_i\rangle = \sqrt{8 m_i T/\pi}$. 
The results presented below are in the kinetic theory limit; we have checked the 
total $\eta$ so obtained matches with that from the Kubo formalism.

In Fig. \ref{visco}(b) we present the shear viscosity $\eta$ for an
equilibrated ensemble from fragmentation of $^{150}$Sm.
At temperatures $T<3$ MeV (corresponding to the liquid phase), the shear viscosity 
is found to rapidly increase with decreasing $T$. For the dilute system comprising of a 
large nucleus and few light particles the mean free path $\lambda$ is large.
Thus a fragment can transport momentum over a large distance resulting in large $\eta$.
At intermediate $T \sim 3-6$ MeV (corresponding to the coexistence region), the
magnitude of viscosity is determined by competing effects between the size and 
multiplicity of the fragments in the system. The Coulomb repulsion will force 
the two colliding nuclei to occupy the available free space (void) which
in turn will collide with the neighboring nucleus and so on. This procedure
can effectively transport momentum over a large distance and produce a large
viscosity \cite{Csernai}; the coefficient grows with temperature as $\eta \sim T^{1/2}$. 
In general, the viscosity $\eta_{A,Z}$ of a species ($A,Z$) was found to progressively 
increase from heavier to lighter particles that have larger $Z/A$ ratios.
Compared to LDM1 at $V=6V_0$, the heavier fragments in LDM2 in the liquid-gas mixed phase 
and somewhat higher temperature (see also Figs. \ref{mulp} and \ref{caloric}) are more 
effective for momentum transport resulting in an increased $\eta$. 
This is due to $\eta_{A,Z}$ values are $\sim 30\%$ higher in the LDM2 
than in LDM1 at a given $T$. At a smaller freeze-out volume $V=3V_0$ fewer 
fragment collisions as well as smaller density of voids inhibits momentum 
transport and thereby reduces dissipation in the medium.

For orientation we also show in Fig. \ref{visco}(b) the shear viscosity from 
analytical fit to the numerical evaluation of $\eta$, obtained by Danielewicz 
\cite{Danielewicz}, in the Uhlenbeck-Uehling transport equations within a first 
order Chapman-Enskog approximation:
\begin{eqnarray} \label{dani}
\eta &=&  (1700/T^2)(n/n_0)^2 + [22/(1+10^{-3}T^2)](n/n_0)^{0.7}  \nonumber\\
&& + 5.8T^{1/2}/(1+160T^2) ~.
\end{eqnarray} 
In this microscopic calculation, where the relevant degrees of freedom are nucleons 
in the colliding nucleus, significant momentum transport by the nucleons
in the early stages of reactions (not considered in our study) leads to faster
growth in viscosity at small $T\sim 1-2$ MeV. In contrast, at $T \gtrsim 4$ MeV, 
the present study clearly underscores the importance of finite size fragments from
multifragmentation (missing in the transport calculations \cite{Danielewicz}) that 
substantially enhance the viscosity in the medium.

\begin{figure}[ht]
\centerline{\epsfig{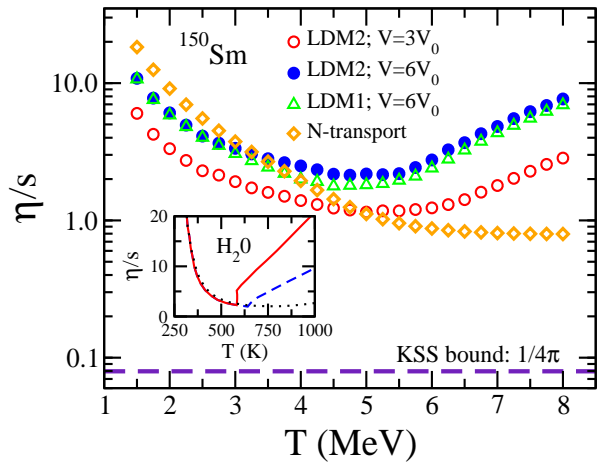}}
\vspace{-0.2cm}

\caption{(Color online) The shear viscosity to entropy density ratio $\eta/s$ 
as a function of temperature $T$ for the breakup of $^{150}$Sm nucleus in LDM1 (triangles) 
and LDM2 (solid circles) mass formula at a freeze-out volume of $V=6V_0$ and in 
LDM2 (open circles) at $V=3V_0$. The $\eta/s$ from nucleon transport is shown in 
diamond. The inset shows $\eta/s$ vs $T$ for water from Ref. \cite{Csernai} 
for an isobar at the critical pressure, $P_c=22.6$ MPa, (dashed lines) and one 
below it at $P=10$ MPa (solid line) and the other above it at $P=100$ MPa (dotted line).}

\label{etas}
\end{figure}

Figure \ref{etas} shows the shear viscosity to entropy density ratio $\eta/s$ 
as a function of temperature in multifragmentation of $^{150}$Sm.
The $\eta/s$ value gradually decreases as a function of rising temperature 
up to $T \sim 4$ MeV in the liquid phase (system dominated by a massive nucleus)
and then increases again at $T \gtrsim 6$ MeV in the gas phase
(system of light particles). At $T \approx 5$ MeV that is
close to the critical temperature for multifragmentation (liquid-gas coexistence phase) 
\cite{Gross,Bondorf,Das,Botvina,Poch} the $\eta/s$ reaches a minimum. 
In the LDM2 mass formula at a freeze-out $V=6V_0$, a minimum of 
$(\eta/s)_{\rm min} \approx 2.1$ is obtained. 
In the absence of surface symmetry energy in LDM1, the $(\eta/s)_{\rm min}$ turns 
to be somewhat smaller. While at a smaller freeze-out volume of $V=3V_0$, the 
$\eta/s$ magnitude is found to be even smaller. The $(\eta/s)_{\rm min}$
estimated in multifragmentation is however significantly above the 
conjectured KSS lower bound of $1/4\pi$ \cite{KSS}. If we adopt the entropy
of the LDM2 set with $V=6V_0$ and $\eta$ of Eq. (\ref{dani}) for nucleon transport 
\cite{Danielewicz}, then $\eta/s$ is seen (Fig. \ref{etas}) to decrease 
continuously with increasing $T$.

Interestingly, the $(\eta/s)_{\rm min}$ obtained in multifragmentation is comparable
to the minimum value for isobars passing in the vicinity of the critical temperature
$T_c$ for liquid-gas phase transition in water \cite{Csernai}. For H$_2$O when 
an isobar passes through the critical point (shown in the inset of Fig. \ref{etas}), 
the $(\eta/s)_{\rm min}$ forms a cusp at $T_c$. When the isobar passes below $P_c$ 
the minimum is at $T<T_c$ with a discontinuous change across the phase transition.
For an isobar passing above $P_c$, a broad minimum is found at a $T$ slightly above $T_c$.
In fact the smallest value of $\eta/s$ corresponds to the most difficult condition
for the transport of momentum. In analogy to this observation, if multifragmentation
is at a fixed freeze-out volume, the hot system will sample a range of $\eta/s$
corresponding to different values of pressures at different temperatures. 
While if the breakup volume varies, the different values of $\eta/s$ found
at $V/V_0=3$ and 6 in Fig. \ref{etas} implies that the system will sample a 
large range of $\eta/s$ in the $(P,V)$ plane to give an average. On the other
hand, if freeze-out is reached at a fixed pressure \cite{Elliot,Chomaz} 
(and close to $P_c$) this would result in a rapid increase in $\eta/s$ at 
$|T-T_c|>0$ comparable to the rise observed for H$_2$O \cite{Csernai}.

In summary, we have studied the thermodynamic and transport properties in fragmentation
of a nucleus within a statistical multifragmentation model. For the equilibrated system 
of fragments and light particles evolving under Coulomb repulsion, we find the shear 
viscosity to entropy density ratio $\eta/s$ exhibits a minimum at a temperature of 
$T \approx 5$ MeV in the coexistence phase of intermediate mass fragments and light particles. 
The minimum value of $(\eta/s)_{\rm min} \simeq 2$ in multifragmentation is comparable to
that at the critical point for liquid-gas phase transition in H$_2$O.
The temperature dependence of $\eta/s$ is somewhat sensitive to the surface effects on 
symmetry energy and depends rather strongly on the freeze-out volume.

\end{document}